\documentclass[journal,transmag]{IEEEtran}
\usepackage{etoolbox}
\makeatletter

\usepackage[round,sort,authoryear]{natbib}

\makeatother
\usepackage{siunitx}
\usepackage{placeins}
\usepackage{url}
\usepackage{textcomp}
\usepackage{diagbox}
\usepackage{fixltx2e}
\usepackage{cite}
\usepackage[cmex10]{amsmath}
\usepackage{amssymb,amsfonts}
\usepackage{algorithmic}
\usepackage{float}
\usepackage{subfigure}
\usepackage{textcomp}
\usepackage[pdftex]{graphicx}
\usepackage{mathabx}
\usepackage{algorithmic}
\usepackage{array}
\usepackage{mdwmath}
\usepackage{mdwtab}
\usepackage{eqparbox}
\newcommand\norm[1]{\lVert#1\rVert}
\begin{document}
% \bstctlcite{IEEEexample:BSTcontrol}
\graphicspath{ {images/} }

\title{Segmentation-guided Domain Adaptation and Data Harmonization of Multi-device Retinal Optical Coherence Tomography using Cycle-Consistent Generative Adversarial Networks\\}

\author{%
    \IEEEauthorblockN{%
        \parbox{\linewidth}{\centering
            Shuo Chen\IEEEauthorrefmark{1},
            Da Ma\IEEEauthorrefmark{2,3,4,1},
            Sieun Lee\IEEEauthorrefmark{5,6},
            Timothy T.L. Yu\IEEEauthorrefmark{1},
            Gavin Xu\IEEEauthorrefmark{1},
            Donghuan Lu\IEEEauthorrefmark{1,7},
            Karteek Popuri\IEEEauthorrefmark{1,8}
            Myeong Jin Ju \IEEEauthorrefmark{9,10},
            Marinko V. Sarunic\IEEEauthorrefmark{1,11,12}, and
            Mirza Faisal Beg\IEEEauthorrefmark{1} \thanks{Corresponding author:  Shuo Chen (shuo\_chen\_4@sfu.ca), Da Ma (dma@wakehealth.edu), Mirza Faisal Beg (faisal-lab@sfu.ca)} % 
        }%
    }%
    \IEEEauthorblockA{%
        \parbox{\linewidth}{\centering
        \IEEEauthorrefmark{1}School of Engineering Science,
        Simon Fraser University, Burnaby, BC, Canada}
    }%
    \IEEEauthorblockA{%
        \parbox{\linewidth}{\centering
        \IEEEauthorrefmark{2} Department of Internal Medicine, Section of Gerontology and Geriatric Medicine,
        Wake Forest University School of Medicine, Winston-Salem, NC, USA}
    }%
    \IEEEauthorblockA{%
        \parbox{\linewidth}{\centering
        \IEEEauthorrefmark{3} Center for Biomedical Informatics, Wake Forest School of Medicine, Winston-Salem, North Carolina, USA}
    }%
    \IEEEauthorblockA{%
        \parbox{\linewidth}{\centering
        \IEEEauthorrefmark{4} Alzheimer's Disease Research Center, Wake Forest School of Medicine, Winston-Salem, North Carolina, USA}
    }%
    \IEEEauthorblockA{%
        \parbox{\linewidth}{\centering
        \IEEEauthorrefmark{5}Mental Health \& Clinical Neuroscience, University of Nottingham, Nottingham, UK}
    }%
    \IEEEauthorblockA{%
        \parbox{\linewidth}{\centering
        \IEEEauthorrefmark{6}Precision Imaging Beacon, University of Nottingham, Nottingham, UK}
    }%
    \IEEEauthorblockA{%
        \parbox{\linewidth}{\centering
        \IEEEauthorrefmark{7}Tencent Jarvis Lab, Shenzhen, China}
    }%
    \IEEEauthorblockA{%
        \parbox{\linewidth}{\centering
        \IEEEauthorrefmark{8} Department of Computer Science, Memorial University of Newfoundland, St. John's, NL, Canada}
    }%
    \IEEEauthorblockA{%
        \parbox{\linewidth}{\centering
        \IEEEauthorrefmark{9} School of Biomedical Engineering, University of British Columbia, BC, Canada}
    }%
        \IEEEauthorblockA{%
        \parbox{\linewidth}{\centering
        \IEEEauthorrefmark{10} Department of Ophthalmology \& Visual Sciences, The University of British Columbia, Vancouver, BC, Canada}
    }%
    \IEEEauthorblockA{%
        \parbox{\linewidth}{\centering
        \IEEEauthorrefmark{11} Institute of Ophthalmology, University College London, London, UK}
    }%
    \IEEEauthorblockA{%
        \parbox{\linewidth}{\centering
        \IEEEauthorrefmark{12} Department of Medical Physics and Biomedical Engineering, University College London, UK}
    }%
}%

\maketitle

\begin{abstract}

% Optical Coherence Tomography(OCT) is a non-invasive technique capturing cross-sectional area of the retina in micro-meter resolutions. It has been widely used as a auxiliary imaging reference to detect eye-related pathology and predict longitudinal progression of the disease characteristics. Retina layer segmentation is one of the crucial feature extraction techniques, where the variations of retinal layer thicknesses and the retinal layer deformation due to the presence of the fluid are highly correlated with multiple epidemic eye diseases like Diabetic Retinopathy(DR) and Age-related Macular Degeneration (AMD). However, these images are acquired from different devices, which have different intensity distribution, or in other words, belong to different imaging domains. This paper proposes a segmentation-guided domain-adaptation method to adapt images from multiple devices into single image domain, where the state-of-art pre-trained segmentation model is available. It avoids the time consumption of manual labelling for the upcoming new dataset and the re-training of the existing network. The semantic consistency and global feature consistency of the network will minimize the hallucination effect that many researchers reported regarding Cycle-Consistent Generative Adversarial Networks(CycleGAN) architecture. 

\textit{Background}: Medical images such as Optical Coherence Tomography (OCT) images acquired from different devices may show significantly different intensity profiles. An automatic segmentation model trained on images from one device may perform poorly when applied to images acquired using a different device, resulting in a lack of generalisability. This study aims to address this issue using domain adaptation methods improved by Cycle-Consistent Generative Adversarial Networks(CycleGAN), especially in cases when the ground-truth labels are only  available in the source domain. 

\textit{Methods}: A two-stage pipeline is proposed to generate segmentation in the target domain. The first stage involves the training of a state-of-the-art segmentation model in the source domain. The second stage aims to adapt the images from the target domain into the source domain. The adapted target domain images are segmented using the model in the first stage. Ablation tests were performed with a different integration of loss functions, and the statistical significance of these models is reported. Both the segmentation performance and the adapted image quality metrics are evaluated. 

\textit{Results}: Regarding the segmentation Dice score, the proposed \textit{ssppg} model achieves a significant improvement of 46.24\% compared to no adaptation, and it reaches 87.4\% of the upper limit of the segmentation performance. Moreover
, the image quality metrics including the FID and KID score indicate that adapted images with better segmentation also have better image qualities. 

\textit{Conclusion}: The proposed method demonstrates the effectiveness of segmentation-driven domain adaptation in retinal imaging processing. It reduces the labour cost of manual labelling, incorporates prior anatomic information to regulate and guide the domain adaptation, and provides insights into improving segmentation qualities in image domains without labels.  

\end{abstract}


\section{Introduction}\label{Intro}
Optical Coherence Tomography(OCT) has become a widely used retinal imaging modality for its benefits of capturing 3D cross-sectional high-resolution retinal structures with non-invasive and cost-effective techniques~\citep{Ikeda2013OpticalTomography}. Both the retinal thickness change and the presence of retinal fluid were biomarkers for various eye diseases. For example, retinal nerve fibre layer (RNFL) thinning was associated with diabetic retinopathy (DR) and glaucoma, and fluid accumulation can indicate diabetic macular edema~\citep{Pekala2019DeepSegmentation}. Deep learning techniques were more robust and accurate than traditional image processing methods for automated segmentation of such clinically important features in retinal OCT images~\citep{Ma2021LF-UNetImages,Roy2017ReLayNet:Networks.}. However, (OCT) images acquired from different devices may show significantly different intensity profiles due to varying properties in the optical systems. Deep neural network (DNN) models trained on a specific dataset might lack generalisability. For instance, a DNN-based segmentation model trained on OCT images acquired from one device using a specific acquisition protocol can experience a significant performance drop when applied to images from different devices and/or protocols. A brute-force solution is to generate manual ground truth data for each device and protocol and re-train the models either locally through the federated-learning framework,~\citep{lo_federated_2021}. However, manual segmentation is costly in terms of time and effort, and it is unfeasible to generate precise manual segmentation for all different devices and scanning protocols. Therefore, a universal and robust automated segmentation algorithm is needed to minimize the labour cost while achieving high segmentation quality across various devices. 

Domain Adaptation (DA) addressed the domain shifting problem by exploring the mapping function between the distributions of the data from the source domain and data from the target domain~\citep{Murez2018ImageAdaptation}. In the context of medical images such as OCT, the concept of "domain" represents a specific device with a specific acquisition protocol with which the data were collected. Unsupervised Domain Adaptation
(UDA) is widely used in scenarios where no labels are available in the target domain, but the images in both domains share certain intrinsic similarities, i.e. the images from either domain can be synthesized from each other~\citep{Toldo2020UnsupervisedReview}. The novel UDA approach, generative adversarial network (GAN), was first introduced in 2014, where the ``adversarial'' minimax game is played between the generative model (generator) and discriminative model (discriminator)~\citep{Goodfellow2014GenerativeNetworks}. The generator tries to adapt the images from the source domain to the target domain, while the discriminator tries to distinguish between the images from the target domain and the generator. The generator and the discriminator are optimized together recursively until both converge. However, the GAN model suffers from several issues. The dynamic equilibrium is difficult to reach due to the unstable oscillations of the two models. The generator may also produce limited variations of synthesized images that can always ``fool'' the discriminator, which is called ``mode collapse''~\citep{Che2017Maximum-LikelihoodNetworks}~\citep{Yan2017MindAdaptation}. On the contrary, the discriminator can become too successful such that the generator learns nothing. Therefore, the constraints including hyper-parameters, loss functions and model complexity should be rigorously designed. Data harmonization, aiming to eliminate the site-specific bias while maintaining the biological characteristics when aggregating multi-site datasets, has proven to be effective through domain adaptation~\citep{Robinson2020Image-levelNetworks}~\citep{Cong2019DoveNet:Verification}. It has been widely used in multi-site multi-scanner MRI datasets with domain variations due to different acquisition protocols~\citep{Tian2022ADataset}~\citep{dinsdale_deep_2021-1}. 

% \subsection{Related work}
Cycle-Consistent Adversarial Networks (CycleGAN) was introduced in 2017, which aims to learn bi-directional mapping functions between the source and target domains~\citep{Zhu2017UnpairedNetworks}. However, the original CycleGAN suffered from the issue of generating hallucination patterns that hinder the successful application in medical and clinical data~\citep{Rahman20213C-GAN:Model}. Several loss functions were proposed to enforce pixel-wise adaptation and cycle consistency of the reconstructed images, and the details will be discussed in section \ref{Methods}. Researchers have dedicated such cross-modality adaptation architecture to various practical applications. Li et al. and Hoffman et al. integrated the semantic consistency concept into the original CycleGAN architecture, which enforces the original image to be segmented consistently as the reconstructed image, and the features of two domains can be distinguished by the discriminator~\citep{Li2019BidirectionalSegmentation}~\citep{Hoffman2018CyCADA:Adaptation}. However, the inaccuracy of the segmentation model may confuse the network from learning the correct structural distribution. Hiasa et al. added the gradient consistency constraint to optimize the segmentation performance near the boundaries of the label for CT-MRI adaptation, where the gradient correlation was a commonly used metric in medical registration that measures the cross-correlation between two images~\citep{Hiasa2018Cross-ModalitySize}. Li et al. adopted a similar strategy by proposing a soft gradient-sensitive loss for the attention of semantic boundaries~\citep{Li2019Semantic-awareAdaption}. The phase consistency in the Fourier domain was studied and proven to be effective by Yang et al. ~\citep{Yang2020PhaseAdaptation}. Zeng et al. further added cross-modality segmentation consistency providing a segmentation model for each modality~\citep{Zeng2021DomainCross-Modality}. These works showed a great potential of modifying CycleGAN with proper constraints in semantic tasks. 

Researchers have also performed several studies regarding the retinal OCT segmentation and domain adaptation using GAN models. He et al. proposed a multi-stage unsupervised domain adaptation network to perform OCT layer segmentation~\citep{He2020SelfNetwork}~\citep{He2020AdversarialSegmentation}. A layer segmentation network (ResUNet) was trained first on Spectralis data, then the auto-encoder was trained on Cirrus images to minimize the segmentation error of the reconstructed images, while the weights of the segmentation network were frozen. The authors observed more retinal surface diffusion and layer boundary shifts using CycleGAN compared to the proposed network. However, the comparison might not be fair as the original CycleGAN paper did not adopt segmentation loss. The observed deformation may not be caused by the changes in the adapted anatomy. In fact, for tasks related to only changed in colour and textures, CycleGAN usually performed well with minimal geometric changes~\citep{Zhu2017UnpairedNetworks}. The segmentation model was sensitive to pixel intensities, especially in OCT images with a low signal-to-noise ratio (SNR) due to speckle noise. Seebock et al. used CycleGAN to adapt Cirrus images to the Spectralis image domain, where a pre-trained UNet model was available to perform retinal fluid segmentation specifically on Spectralis images~\citep{Seebock2019UsingSegmentation}. The segmentation performance on the adapted images was comparable to the pre-trained segmentation model. However, both of these works did not reveal the potential of GAN due to the limited number of data, unconsolidated constraints of loss functions, etc. 

In this paper, we propose a segmentation-guided CycleGAN network aiming to achieve the universal retinal OCT segmentation model. The experiment set assumes that images from the source domain are fully labelled, while the images from the target domain are not labelled. This study has the following contributions:
\begin{enumerate}
    \item We propose a novel two-staged CycleGAN-based network designed for layer segmentation of retinal OCT images acquired from different devices
    \item We incorporate both the ground-truth segmentation information and the pre-trained segmentation model from the source domain to guide the CycleGAN-based domain adaptation.
    \item Our approach reduces the hallucination effect that the original CycleGAN network suffered from, and we obtain a segmentation result comparable to the model trained on ground-truth data.
\end{enumerate}

\section{Methods}\label{Methods}

\subsection{Data Acquisition}\label{dataAcq}
We adopted two datasets for training and validation purposes. In the first cohort (source domain), 60 OCT volumes of 60 patients acquired from the Zeiss Cirrus 5000 HD-OCT machine (Zeiss Meditec. Inc, Germany) were collected from Vancouver Northshore Clinic. Each volume contained 245 Bscans with a pixel dimension of 1024x245. They were manually labelled with 5 retinal layers, 2 background regions including Vitreous and Choroid-Sclera complex, and fluid. For simplification, the Choroid-Sclera complex will be referred to as Choroid in the following content. More details about the dataset, acquisition protocol and manual segmentation protocol can be found in~~\citep{Ma2021LF-UNetImages}. In the second cohort (target domain), 10 OCT volumes of 60 different patients acquired from Topcon 3D OCT 1000 Mk2 machine (Topcon. Inc, Japan) were downloaded from the UK Biobank database and manually labelled with 8 retinal layers~\citep{SirRoryCollinsUKBiobank}. Each Topcon volume contained 128 Bscans with a resolution of 650x512. All Topcon data were acquired from the healthy patients, i.e. there was no fluid present. We will refer to these two datasets as the Zeiss dataset and Topcon dataset, respectively.

\begin{figure}[htb]
\centering
    \subfigure[Example of the segmentation of 5 inner retinal layers for Zeiss OCT Bscan]{
    \includegraphics[width=0.45\textwidth]{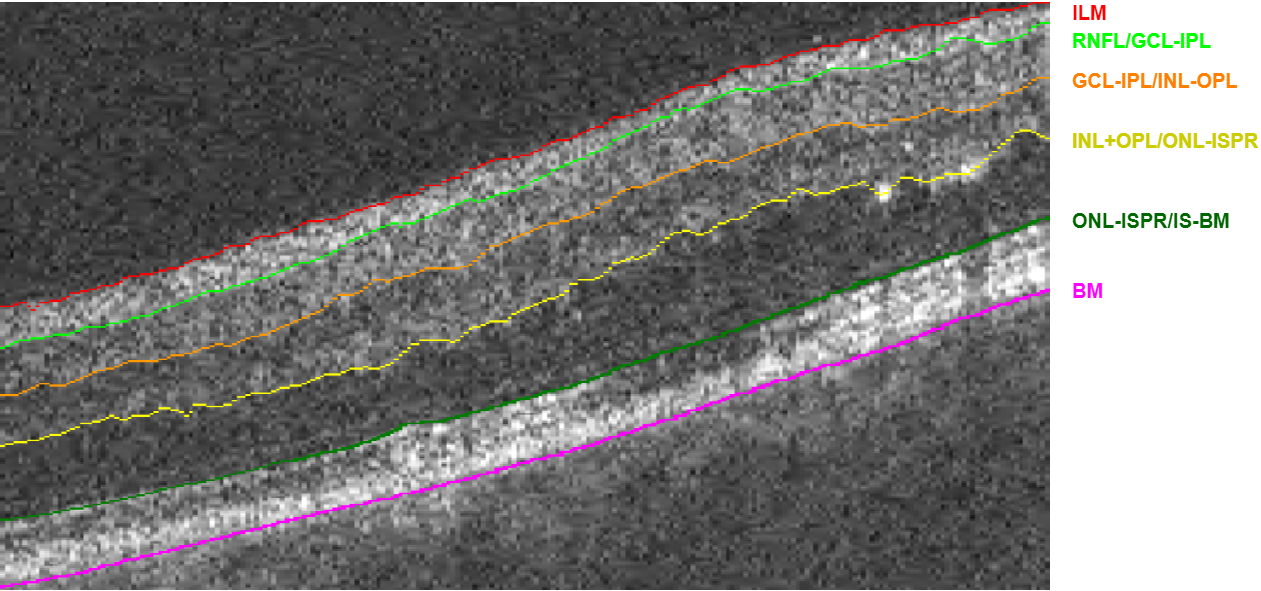}}
    \quad
    \subfigure[Example of the segmentation of 5 inner retinal layers for Topcon OCT Bscan]{
    \includegraphics[width=0.45\textwidth]{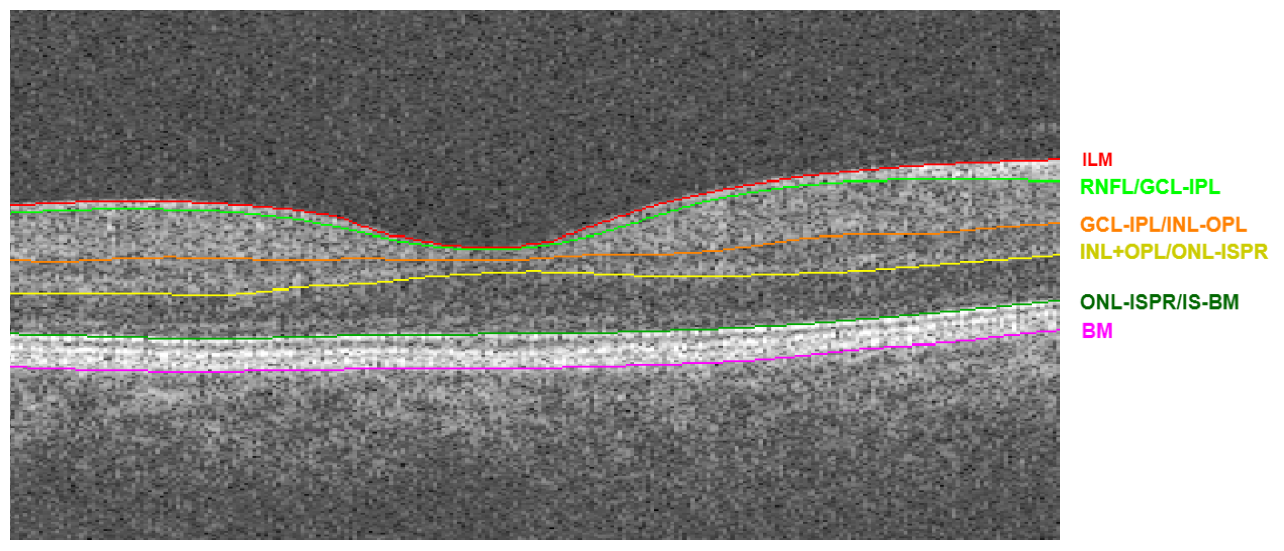}
    }
\caption{Demonstration of the segmentation of 5 inner retinal layers for the Zeiss dataset and Topcon dataset, respectively. The corresponding layer boundaries are highlighted and categorized as shown on the right side of the image.}
\label{fig:layerseg}
\end{figure}

Figure \ref{fig:layerseg} demonstrates examples of the manual segmentation ground truth labels for both the Zeiss dataset and Topcon dataset, respectively. For both datasets, seven regions are defined and labelled (five retinal layers plus two non-reginal regions). The five retinal layers from top to bottom are ILM(Inner Limiting Membrane)-RNFL(Retinal Nerve Fiber Layer), GCL(Ganglion Cell Layer)-IPL(Inner Plexiform Layer) complex, INL(Inner Nuclear Layer)-OPL(Outer Plexiform Layer) complex, ONL(Outer Nuclear Layer)-ISPR(Inner-Segment Photoreceptor Layer) complex, IS(Inner-Segment Layer)-BM(Bruch's Membrane). In addition, the region above ILM (Internal Limiting Membrane) is considered Vitreous, and the region below BM(Bruch's Membrane) is regarded as Choroid. The region above ILM is considered Vitreous, and the region below BM is regarded as Choroid. Each colored trajectory is defined by the boundaries of layers, e.g. RNFL/GCL-IPL means the boundary between the RNFL layer and the GCL-IPL complex. % The Topcon dataset was initially segmented into 10 retinal regions, where 3 extra inner retinal layers are delineated. For consistency, we re-arrange them to 7 regions equivalent to the Zeiss dataset. 

% Figure \ref{fig:layerseg} demonstrates examples of the 8-layer manual segmentation for the Topcon dataset and 5-layer manual segmentation for the Zeiss dataset, respectively. The Topcon dataset was segmented into 8 retinal layers. The inner retinal layers from top to bottom are RNFL (Retinal Nerve Fiber Layer), GCL (Ganglion Cell Layer)-IPL(Inner Plexiform Layer) complex, INL (Inner Nuclear Layer), OPL (Outer Plexiform Layer), ONL (Outer Nuclear Layer)-ISPR (Inner-Segment Photoreceptor Layer) complex, IS-OS (Inner Segment-Outer Segment/Myoid Zone) junction, OSPR1 (Outer Segment Photoreceptor Layer 1/Ellipsoid Zone) and OSPR2 (Outer Segment Photoreceptor Layer 2/OS-RPE complex). The region above ILM (Internal Limiting Membrane) is considered Vitreous, and the region below BM (Bruch's Membrane) is regarded as Choroid. For the Zeiss dataset, 7 regions are defined and labelled instead. Specifically, besides the same Vitreous and Choroid region, the inner retinal regions from top to bottom are ILM\_NFL(RNFL), NFL\_IPL(GCL-IPL complex), IPL\_OPL(INL+OPL), OPL\_IOS(ONL-ISPR complex + IS-OS junction), IOS\_BM(OSPR1+OSPR2), respectively. Each region is defined by the boundaries of layers, e.g. ILM\_NFL means the region from the lower boundary of ILM to the upper boundary of NFL. The layers within the square brackets show the equivalent regions defined in Topcon settings. In other words, we rearranged the 8-layer UKB ground-truth labels to the Zeiss standard labels with 5 inner retinal layers.

We applied two common pre-processing steps to optimize the performance~\citep{Ma2022ClinicalLearning.,lee_age_2017}. Motion correction was performed using phase cross-correlation between adjacent Bscans to calculate both the relative axial and lateral shifts with the first Bscan as a reference, and then 2D spine interpolation was applied to obtain the motion-corrected Bscans. Bounded variation (BV) 3D smoothing was applied to minimize the effect of speckle noise and enhance the contrast of the retinal layer and fluid boundaries. For both Zeiss and Topcon datasets, we cropped the excessive background regions to obtain a Bscan size of 500x231 and 256x512, respectively. It helped both the segmentation and adaptation network to learn more about the retinal layers inside the retinal body. We resized the Bscans from both the Zeiss dataset and the Topcon dataset to a resolution of 512x256 before feeding them into the domain adaptation network.

\subsection{Network}\label{network}
The overall pipeline contains two steps, the segmentation network and the device domain adaptation network. The segmentation network was trained first for optimal performance of supervision of the adaptation network. Then, the CycleGAN-based adaptation network was trained while the weight of the segmentation network was frozen. The segmentation loss along with several other auxiliary constraints were combined throughout the training. 
\subsubsection{OCT Segmentation Network}\label{network_seg}
\begin{figure*}[htb]
\centering
\includegraphics[width=0.7\textwidth]{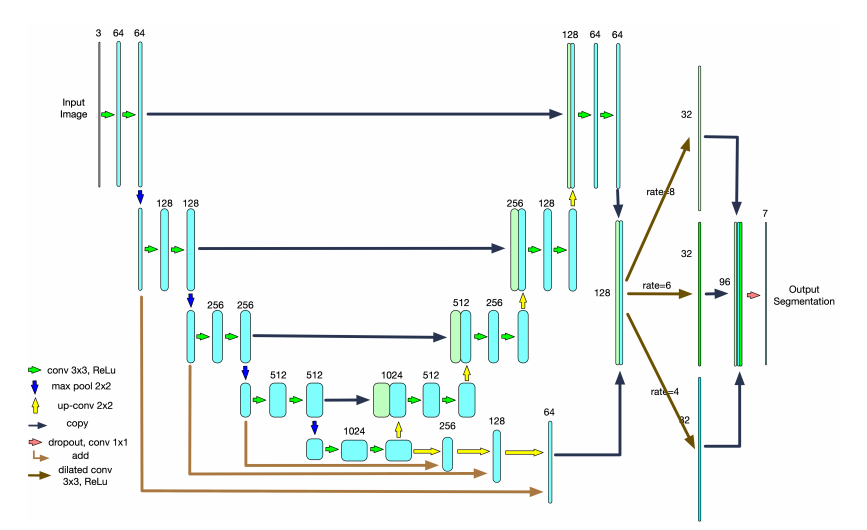}
\caption{Illustration of the LF-UNet architecture. It mimics the encoder-decoder design of the original U-Net structure, where the contracting path encoded the high-level features, and the expanding path decoded these features to reconstruct the corresponding semantic information. An extra fully-connected path is added alongside the expanding path to better increase the segmentation accuracy of the retinal layer boundaries. More details are discussed by Ma et al.~\citep{Ma2021LF-UNetImages}}.
\label{fig:LFUnet}
\end{figure*}

The LF-Unet, a deep neural network integrating U-Net and fully convolutional network(FCN), was used as a pre-trained OCT segmentation network ~\citep{Ma2021LF-UNetImages}. As shown in Figure \ref{fig:LFUnet}, the architecture of the LF-UNet leveraged a symmetrical convolutional network with an encoder-decoder mechanism. Mimicking the original UNet structure, the contracting paths contain 4 down-sampling convolutional blocks to extract high-level features. 2 expansive paths include the original up-sampling path and an FCN path. The output was eventually fed into 3 consecutive dilated convolutional layers for final results.

The network was trained on the Zeiss dataset using the PyTorch Lightning framework with 2-node distributed learning~\citep{WilliamFalconPyTorchLightning}. We set the batch size and learning rate to be 2 and $10^{-4}$, respectively. We used Adam optimizer with L2 regularization. We applied early stopping criteria with patience of 3 so that the training was terminated when the validation loss converges. We used the learning scheduler to monitor the validation loss and reduce the learning rate when validation loss plateaued ('ReduceLROnPlateau' command in PyTorch Lightning). We applied several data augmentations by random horizontal flipping and random spatial rotation with a maximum angle of 10 degrees. We applied a pixel-wise Softmax function for the model output and uses a weighted combination of Dice loss and Cross Entropy Loss to enforce the learning of the fluid under the class imbalance circumstance. We calculated the pixel-wise weight map for each Bscan by assigning more weights to the fluid region and boundary. Given predicted segmentation $S_p$, ground-truth segmentation $S_g$ and pixel-wise dice weights $S_{dw}$, class weights $S_{cw}$ the segmentation loss is calculated as:
\begin{equation}
L_{dice} =  1 - \frac{\sum 2S_{p}S_{g}S_{dw}}{\sum S_{p}S_{dw}\sum S_{g}S_{dw}  + \epsilon}
\end{equation}

\begin{equation}
L_{ce} = -\sum S_{cw}S_{g}log(S_{p})
\end{equation}

\begin{equation}
L_{seg} = L_{dice} + L_{ce}
\end{equation}

To avoid over-fitting and ensure the robustness of the segmentation network performance, 10-fold validation was applied for the entire dataset. Each fold was split into training, validation and testing sets with a ratio of 8:1:1. We applied the volume stratification during the splitting such that the Bscans from any OCT volume were allocated to one set only. It avoided biased evaluation when adjacent Bscans with similar features from the same OCT volume are exposed to multiple training stages. We evaluated the performance of each fold of the model with the average dice score using their corresponding test set. The model with the best average dice score was adopted for the next stage of training of the adaptation network.

\subsubsection{Domain Adaptation Network}\label{network_adapt}
\begin{figure*}[htb]
\centering
\includegraphics[width=0.6\textwidth]{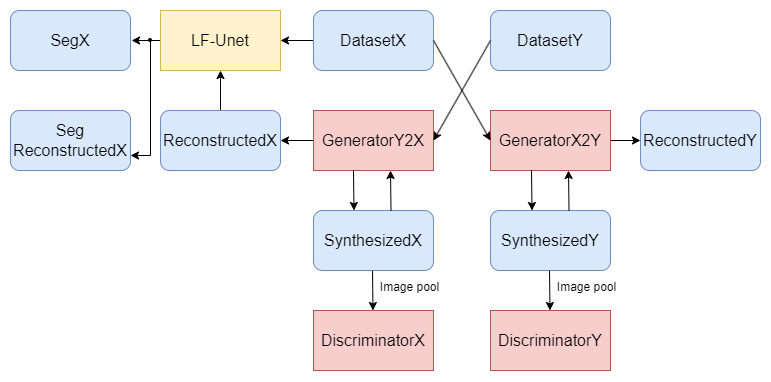}
\caption{Workflow for the domain adaptation network. The images from DatasetX and DatasetY are fed into the GeneratorX2Y and GeneratorB2A, respectively. The GeneratorX2Y and GeneratorY2X can then produce adapted images of SynthesizedY and SynthesizedX, respectively. A series of SynthesizedX and SynthesizedY stored in image pools are then fed into the DiscriminatorX and DiscriminatorY accordingly for GAN loss calculations. Meanwhile, the SynthesizedX and SynthesizedY are fed back to the GeneratorX2Y and GeneratorY2X to produce their corresponding reconstructed images RecontstructedY and RecontstructedX, respectively. The images from DatasetX and ReconstructedX images are used by LFUnet to generate the segmentation SegX and SegReconstructedX, which are utilized for semantic loss calculation. }
\label{fig:flow}
\end{figure*}

Similar to the original CycleGAN architecture, the segmentation-guided adaptation network consists of two generators and two discriminators. As shown in Figure \ref{fig:flow}, datasets A and B refer to the Zeiss and Topcon dataset, respectively. Image pairs from both datasets were fed into the corresponding generators simultaneously. The output synthetic adapted image pairs were then fed into two branches. The discriminators took a history of the synthesized images to distinguish if the upcoming pairs were real or synthetic. Meanwhile, the reconstructed images were obtained by generators fed with the adapted images. Lastly, the original and reconstructed image pairs were fed into the pre-trained segmentation network to obtain the predicted segmentation pairs. 

\begin{figure*}[htb]
\centering
\includegraphics[width=0.7\textwidth]{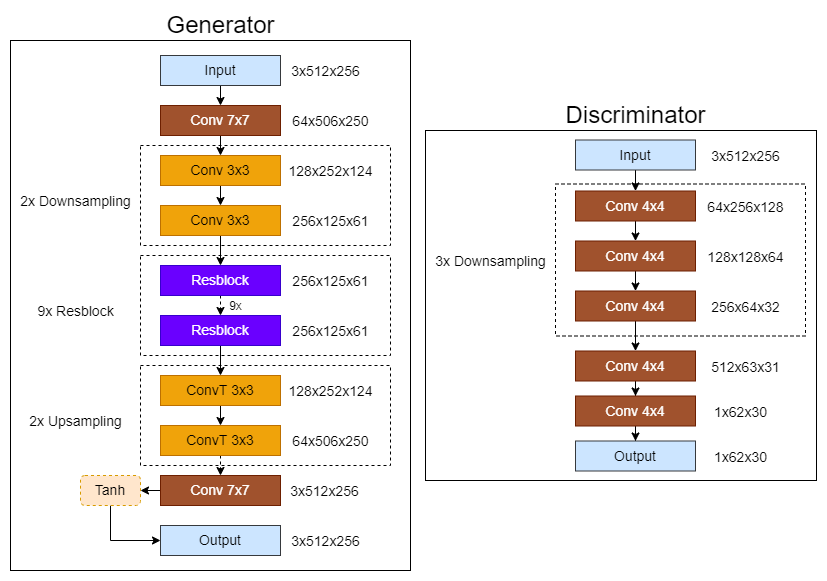}
\caption{Generator and Discriminator architecture. The dimension of the output intermediate feature map is shown on the right of each block, which is formatted as $Channel \times Height \times Weight$. The Generator block is implemented as an encoder-decoder architecture. 9 Resblocks are placed at the bottleneck for high-level feature extractions. The Discriminator is implemented with a series of down-sampling convolutions, which ends up with a single-channel classification map filled with probabilities of images being real(1) or synthesized(0). Each element in the classification matrix represents a local region in the original image, which is originally proposed in PatchGAN~\citep{Isola2017Image-to-imageNetworks}}
\label{fig:G&D}
\end{figure*}

As shown in Fig \ref{fig:G&D}, the generator block was designed with an encoder-decoder mechanism. We used 2-level convolutional down-sampling to obtain high-level features. It was then fed into 9 consecutive residual blocks(Resblock) to learn cross-domain correlations. Lastly, the adapted latent features were recovered to the original size with Tanh activation. The discriminator block was constructed similarly to VGG16 architecture. Multiple convolutional layers were used for high-level feature discrimination. A single-channel classification map was generated as output.

\begin{figure}[htb]
\centering
\includegraphics[width=0.45\textwidth]{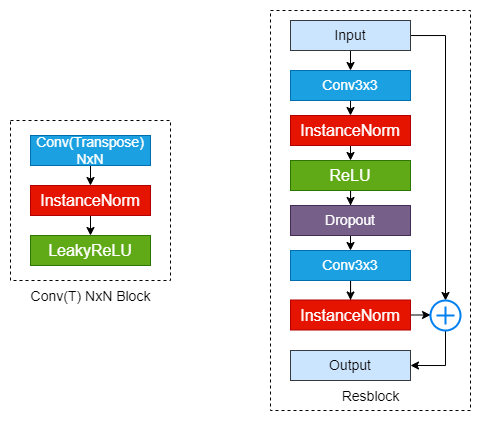}
\caption{Convolution block and Resblock used in both Generator and Discriminator blocks. The Conv NxN block is a 3-step operation with the 2D convolution of input size NxN, followed by normalization and LeakyReLU activation layers. The up-sampling operation will replace the standard convolution with transposed convolution. The Resblock is constructed with two consecutive Conv NxN blocks with ReLU activation and a Dropout layer in between.}
\label{fig:blocks}
\end{figure}

As shown in Fig \ref{fig:blocks}, we used the LeakyReLU activation function to avoid dying neuron problems for ReLU. To cooperate with this, all the input images were normalized within the range of [-1,1], but they were demonstrated with a re-scaling back to [0,255]. Also, we adopted a dropout layer in Resblock to reduce the effect of over-fitting. 

Despite the difference in capacities between the Zeiss and Topcon dataset, we formed the image pairs for adaptation using all available data. For the Zeiss dataset, we adopted the same data split used for the training of the selected segmentation model. For the Topcon dataset, we fixed one volume as a hold-out test set, and perform 9-fold inner cross-validation along with the training and validation dataset from the Zeiss dataset. Specifically, during each fold of training, we split the 9 Topcon volumes with a ratio of 8:1 for training and validation, respectively. Thus, each image pair was formed from Bscans of 6 Zeiss volumes and 1 Topcon volume. For reproducible results, we fixed the combination of volumes and Bscans for image pair formation. 

We applied several techniques to optimize the performance of the adaptation network. We applied the same data augmentation as the segmentation network. Since we wanted to adapt the images directly acquired from the OCT devices with few processing steps as mentioned in Section \ref{dataAcq}, the intensity-level augmentation may confuse the network of constructing stabilized mapping function. 
We adopt the tricks used in WassersteinGAN from Arjovsky \textit{et al.} to use RMSProp optimizer for parameter adjustment~\citep{Arjovsky2017WassersteinGAN}. The gradient changes for GAN were usually unstable, so the optimizer using a momentum mechanism may cause gradient clipping. We set the initial learning rate of the discriminator to be 5 times larger than the generator according to Two time-scale update rule (TTUR), which helped the convergence to a local Nash equilibrium and works functionally equivalent to train the discriminator more frequently~\citep{Heusel2017GANsEquilibrium}. For losses related to the discriminator, we applied soft/noisy labels with a pixel-wise variation of a normal distribution drawn between [-0.2, 0.2], which helped enhance the generalisability of the whole network. We set a learning rate of $10^{-5}$ to stabilize the adaptation. The model was trained with a maximum of 100 epochs while the learning rate is linearly decayed after 50 epochs. We applied early stopping criteria that the training stops when the validation segmentation loss of the UKB dataset stops decreasing after 3 epochs. We used the PyTorch Lightning Framework for 2-node distributed parallel training with a batch size of 2. 

\subsection{Domain Adaptation Loss functions}\label{network_loss}

\subsubsection{CycleGAN Losses}
The original CycleGAN paper mentioned three loss functions. GAN loss, or adversarial loss, was designed to evaluate the performance of the discriminator fed with synthetic data. Given the images x in the source domain X (Zeiss dataset) and images y in the target domain Y (Topcon dataset), the generator functions $G_{X2Y}$ and $G_{Y2X}$ for mapping from X to Y and Y to X, respectively, and discriminator functions $D_{X}$ and $D_{Y}$, we have:

\begin{multline}
    L_{GAN}(x,y) = MSE(D_Y(G_{X2Y}(x)), 0) + \\ MSE(D_X(G_{Y2X}(y)), 0)
\end{multline}

\begin{equation*}
    MSE(x,y) = (x - y)^2
\end{equation*}

The adversarial loss can only guide the network with style transfer across domains, but it cannot guarantee the unique mapping of the two domains. Cycle consistency loss was then proposed that the synthetic images should also be reconstructed back to the original images:

\begin{multline}
L_{cyc}(x,y) = \lambda_X \norm{x - G_{Y2X}(G_{X2Y}(x))} + \\ \lambda_Y \norm{y - G_{X2Y}(G_{Y2X}(y))}
\end{multline}

Here, the $\lambda_X$ and $\lambda_Y$ are tuning parameters from both adapted directions.

To further emphasize the independence of the two adaptation functions, the identity loss was designed such that the original image should remain unchanged if the adaptation function in a reversed direction was applied:

\begin{equation}
L_{idt}(x,y) = \norm{x - G_{Y2X}(x)} + \norm{y - G_{X2Y}(y)}
\end{equation}

\subsubsection{Segmentation Loss}
Even though the above loss functions proposed in the original CycleGAN enforced a unique mapping between the two domains, the segmentation model was not guaranteed to perform well onto a such particular solution. As described in Section \ref{network_seg}, we had a pre-trained segmentation model on the Zeiss dataset. Thus, the semantic consistency loss was designed such that the segmentation model should have comparable performance on the original image x and its reconstructed image. Given the ground-truth labels l for the original image, the segmentation function S, and the same segmentation criterion $L_{seg}$ as Eq. (3), we have:

\begin{multline}
L_{sem}(x, l) = L_{seg}(S(x), l) + \\ L_{seg}(S(G_{Y2X}(G_{X2Y}(x))), l)
\end{multline}

Since the weights of the segmentation network were frozen during the training of the adaptation network, the first term of the semantic consistency loss should remain relatively stable, and it performed as a ``self-awareness'' factor of the error caused by the segmentation network itself. The second term will further guide the cyclic adaptation to achieve comparable segmentation.

\subsubsection{Structure Similarity Loss}
Wang et al. proposed Structure Similarity Index Measure (SSIM) to quantify the similarity between two images in luminance, contrast and structure~\citep{Wang2004ImageSimilarity}. Luminance was measured based on the average intensity value, contrast was measured by the standard deviation of the intensity, and structure was calculated based on the normalized correlation between the two images. Based on SSIM, we defined a difference SSIM (DSSIM) loss to maximize the exponentially weighted multiplication of these three metrics between the original image and its reconstruction. Given the mean \(\mu_x\) \(\mu_y\), and standard deviation \(\sigma_x\) \(\sigma_y\) for images x and y, respectively, we have:

\begin{equation}
L_{dssim}(x, y) = (1 - \frac{(2\mu_x\mu_y+c_1)(2\sigma_{xy}+c_2)}
      {(\mu_x^2+\mu_y^2+c_1)(\sigma_x^2+\sigma_y^2+c_2)}) / 2
\end{equation}

This expression was obtained by setting the weights for all three losses evenly, and $c_1$ and $c_2$ were constants included to prevent the denominator terms from being close to 0.

\subsubsection{Peak Signal to Noise Ratio Loss}
OCT images suffers from speckle noise due to the spatially coherent light source~\citep{Bashkansky2000StatisticsTomography}. The Peak Signal to Noise Ratio (PSNR) can be characteristic of images from specific OCT devices providing the images were generated consistently. The segmentation model is usually sensitive to the intensity distribution of the image. Therefore, the different PSNR will directly affect the segmentation quality. The PSNR loss was constructed such that the images from the same domain should share the same noise distribution:

\begin{equation}
L_{psnr}(x, y) = - 10 \log_{10} \frac{max^2(x)}{MSE(x,y)}
\end{equation}

\subsubsection{Perceptual Loss}
The training of a GAN is time-consuming and difficult to optimize due to the obstacles mentioned in \ref{Intro} Introduction. Johnson et al. provided a novel approach that the high-level features of an image should not be lost during adaptation~\citep{Johnson2016PerceptualSR}. In other words, the original image, synthetic image and reconstructed image are supposed to share structural similarity in high-level convolutional feature extraction layers. In our training, four convolutional layers of a pre-trained VGG16 network from ImageNet were used to extract each level of feature~\citep{Simonyan2014VeryRecognition}. It speeded up the converging process by reconstructing the high-resolution image directly from low-resolution features. Given the output of the $i^{th}$ layer of the VGG16 network $f_i(x)$ when fed with image x, and $f_i(y)$ when fed with image y, the size of both input images is (C,H,W), and the perceptual loss is defined as:

\begin{multline}
L_{perc}(x, y) = \frac{1}{C \times H \times W} \times \sum_{i=1}^{4} (\norm{f_i(x) - f_i(G_{X2Y}(x))} + \\  \norm{f_i(y) - f_i(G_{Y2X}(y))} + \norm{f_i(x) - f_i(G_{Y2X}(G_{X2Y}(x)))} + \\ \norm{f_i(y) - f_i(G_{X2Y}(G_{Y2X}(y)))} + \\ \norm{f_i(G_{X2Y}(x)) - f_i(G_{Y2X}(G_{X2Y}(x)))} + \\ \norm{f_i(G_{X2Y}(y)) - f_i(G_{X2Y}(G_{Y2X}(y)))})
\end{multline}

Instead of the euclidean distance used in the original paper, we adopted L1 norm for robustness of overall adaptation, since it was common that some low-quality OCT images exist.

\subsubsection{Gradient Consistency Loss}
To further maintain the texture of the source images, especially the layer boundaries where the segmentation network focused on, we proposed gradient consistency loss that enforced a better adaptation of regions with large intensity variations. The assumption was that the adaptation shall not diminish the visually-recognizable edges, especially the boundaries of retinal layers. The edge information was extracted and compared using 1st order image derivative in both x and y direction with 2D Sobel operator:

\vspace{5mm}

\begin{equation*}
G_x = \begin{pmatrix} -1 & 0 & 1\\-2 & 0 & 2\\-1 & 0 & 1\end{pmatrix},
G_y = \begin{pmatrix} 1 & 2 & 1\\0 & 0 & 0\\-1 & -2 & -1\end{pmatrix}
\end{equation*}

\vspace{5mm}

Unlike the approach of Li et al. where the attention is emphasized mostly on layer boundaries, we simply used the intrinsic information of the images to maintain more useful information. Let * denote the convolution operation, the gradient consistency loss is constructed as:

\begin{multline}
L_{grad}(x, y) = MSE(G_x \ast x, G_x \ast G_{X2Y}(x)) + \\ MSE(G_y \ast y, G_y \ast G_{Y2X}(y)) \\
\end{multline} 

Thus, the overall loss function for the adaptation network is:

\begin{multline}
L_{tot} = L_{GAN} + L_{cyc} + \lambda_{idt} L_{idt} + \\ \lambda_{seg} L_{sem}+ \lambda_{dssim} L_{dssim}+ \lambda_{psnr} L_{psnr} + \\ \lambda_{feat} L_{perc} + \lambda_{grad} L_{grad} 
\end{multline}

Here, the $\lambda_{idt}$, $\lambda_{seg}$, $\lambda_{feat}$, $\lambda_{dssim}$, $\lambda_{psnr}$, $\lambda_{grad}$ along with $\lambda_X$ and $\lambda_Y$ mentioned before are tuning hyper-parameters for each corresponding loss term. For all the experiments, we empirically set $\lambda_{X}$ and $\lambda_{Y}$ to be 20 for cycle consistency loss, 0.5 for $\lambda_{idt}$, and all other tuning parameters to be 1. 

\subsection{Domain Adaptation Model Evaluation}

We evaluated the performance of the domain adaptation on the test sets from the target domain (i.e. domain Y of the Topcon images). Both the segmentation-based and the image-based metrics were evaluated to achieve a comprehensive evaluation of the domain adaptation results that cover task applications.

\subsubsection{Segmentation-based metrics evaluation}\label{seg_metrics}

Firstly, the performance of the segmentation network trained on the source domain (Zeiss) was evaluated by computing the segmentation accuracy using the Dice similarity coefficient, or dice score (Equation \ref{eq:dice}) for each retinal layer labels, between the ground truth manual segmentation in the test set and the automatic segmentation trained from the LF-UNet. 

\begin{equation}
    Dice = \frac{2(X\bigcap Y)}{|X|+|Y|}
    \label{eq:dice}
\end{equation}
where X is the ground truth label maps generated from manual segmentation, and Y is the automatically-generated label maps generated from the proposed whole-body multi-slice segmentation framework.

Secondly, the performance of domain adaptation was further evaluated using each retinal layer label. Dice scores between the auto-segmentation results derived from the first-level segmentation model (LF-UNet) and the automatic segmentation results derived from the domain-adapted images in the source domain were compared. 

\subsubsection{Image-based metrics evaluation}\label{image_metrics}

Other than directly evaluating the segmentation performance, we also evaluated the quality of the adapted images, where the variations may not be visible to human eyes. Heusel et al. first proposed the Fréchet Inception Distance (FID) score to evaluate the similarities between two datasets, which was widely used for measuring the performance of GAN via computing the distance between two multidimensional Gaussian distributions~\citep{Heusel2017GANsEquilibrium}. It adopted the Inception V3 as a feature extractor and computes the Fréchet distance between the high-level feature maps~\citep{Dowson1982TheDistributions}~\citep{SzegedyRethinkingVision}. Given the original and the adapted distribution to be X and Y, $\mu_X$ and $\mu_Y$ to be the mean of the two distributions,  $\Sigma_X$ and $\Sigma_Y$ to be the covariance matrices of the two distributions, $Tr$ representing the trace of the matrix, the FID score is formulated as:

\begin{multline}
FID(X,Y) = \norm{\mu_X - \mu_Y}^2 - \\ Tr(\Sigma_X + \Sigma_Y - 2\Sigma_X \Sigma_Y)
\end{multline} 

In addition, Gretton et al. proposed a kernel-based method to evaluate the similarity of two-sample distribution using Maximum Mean Discrepancy (MMD)~\citep{Gretton2012ATest}. It assumed that two different distributions shall possess different expected values, which can then be utilized to distinguish different empirical datasets~\citep{Borgwardt2006IntegratingDiscrepancy}. It maps the L2 distance of two distributions into a universal Reproducing Hilbert Kernel Space (RHKS), providing that the mapping function is a unit ball~\citep{Borgwardt2006IntegratingDiscrepancy}. The MMD score requireed the calculation of a polynomial kernel function. Given two data samples x and y, scaling coefficients $\gamma$ and $c$, degree of the polynomial kernel $n$, the polynomial kernel $k(x,y)$ are defined as:

\begin{equation*}
k(x, y) = (\gamma \cdot x^Ty+c)^n
\end{equation*}

Then, given two data distributions X and Y, the number of samples m and n, for all $x_{i} \in X, y_{i} \in Y$, the MMD score is calculated as:

\begin{multline*}
MMD = [\frac{1}{m(m-1)} \sum_{i \neq j}^{m} k(x_i, x_j) \\ + \frac{1}{n(n-1)} \sum_{i \neq j}^{m} k(y_i, y_j) - \frac{2}{mn} \sum_{i,j=1}^{m,n} k(x_i, y_j)]^\frac{1}{2}
\end{multline*}

We calculated the mean and standard deviation using the Kernel Inception Distance (KID) via MMD score, which is calculated as:

\begin{equation}
KID(X, Y) = MMD(X, Y)^2
\end{equation}

\section{Results}\label{results}
We obtained the segmentation results for both stages of the pipeline, i.e. the segmentation network and adaptation network. We also evaluated the quality of the adapted images compared to the original images within the same domain. To maintain consistency with the adaptation network, we evaluated the similarities in the domain of the Zeiss dataset via commonly used metrics mentioned in Section \ref{image_metrics}.

\subsection{Segmentation network}\label{results_seg}
For the segmentation network mentioned in Section \ref{network_seg}, we applied 10-fold cross-validation to evaluate the segmentation performance. For each fold, the best model was selected with the lowest validation loss. The average dice score was then computed by feeding the corresponding test set data into the selected model. The average dice scores for 8 retinal regions are shown in Table \ref{tbl:dice_seg}. The fold 2 model has the highest average dice score among all folds, thus it is used for the next stage of the adaptation network. Notice that the dice score of fluid was also considered as we wanted to minimize the false positive rate of fluid when the segmentation network was applied to Topcon dataset with only healthy patients.

\begin{table*}[htb]
\centering
\begin{tabular}{|c|c|c|c|c|c|c|c|c|c|}
\hline
\diagbox{Fold}{Dice}{Layer} & $ILM\_NFL$ & $NFL\_IPL$ & $IPL\_OPL$ & $OPL\_IOS$ & $IOS\_BM$ & $Fluid$ & $Vitreous$ & $Choroid$ & $Average$\\
\hline
1 & 0.8832 & 0.9354 & 0.9408 & 0.9524 & 0.9329 & 0.7706 & 0.9963 & 0.9908 & 0.9253\\
\hline
2 & \textbf{0.9044} & \textbf{0.9629} & 0.9456 & 0.9546 & 0.9477 & \textbf{0.8342} & 0.9904 & 0.9918 & \textbf{0.9414}\\
\hline
3 & 0.9007 & 0.9622 & \textbf{0.9541} & \textbf{0.9668} & \textbf{0.9550} & 0.6506 & 0.9975 & 0.9939 & 0.9226\\
\hline
4 & 0.9008 & 0.9621 & 0.9466 & 0.9514 & 0.9449 & 0.6796 & \textbf{0.9977} & 0.9922 & 0.9219\\
\hline
5 & 0.8786 & 0.9390 & 0.9242 & 0.9564 & 0.9392 & 0.6097 & 0.9974 & 0.9911 & 0.9045\\
\hline
6 & 0.8941 & 0.9449 & 0.9366 & 0.9480 & 0.9226 & 0.7005 & 0.9974 & 0.9878 & 0.9165\\
\hline
7 & 0.8885 & 0.9387 & 0.9104 & 0.9010 & 0.9052 & 0.3764 & 0.9956 & 0.9873 & 0.8629\\
\hline
8 & 0.9042 & 0.9509 & 0.9208 & 0.9216 & 0.9354 & 0.4365 & 0.9969 & 0.9910 & 0.8821\\
\hline
9 & 0.8853 & 0.9446 & 0.9296 & 0.9299 & 0.9433 & 0.5356 & 0.9968 & 0.9919 & 0.8946\\
\hline
10 & 0.8932 & 0.9296 & 0.9219 & 0.9538 & 0.9533 & 0.5996 & 0.9969 & \textbf{0.9942} & 0.9053\\
\hline
\end{tabular}
\caption{Mean dice score of 7 retinal regions and fluid for the different fold of validations. The best model is selected based on the average dice scores among all regions. The best dice score for each retinal region is highlighted. Fold 2 is chosen with the highest average dice score of 0.9414}
\label{tbl:dice_seg}
\end{table*}

\subsection{Adaptation network}\label{results_adapt}
\begin{figure}[htb]
\centering
\includegraphics[width=0.45\textwidth, height=0.7\textheight]{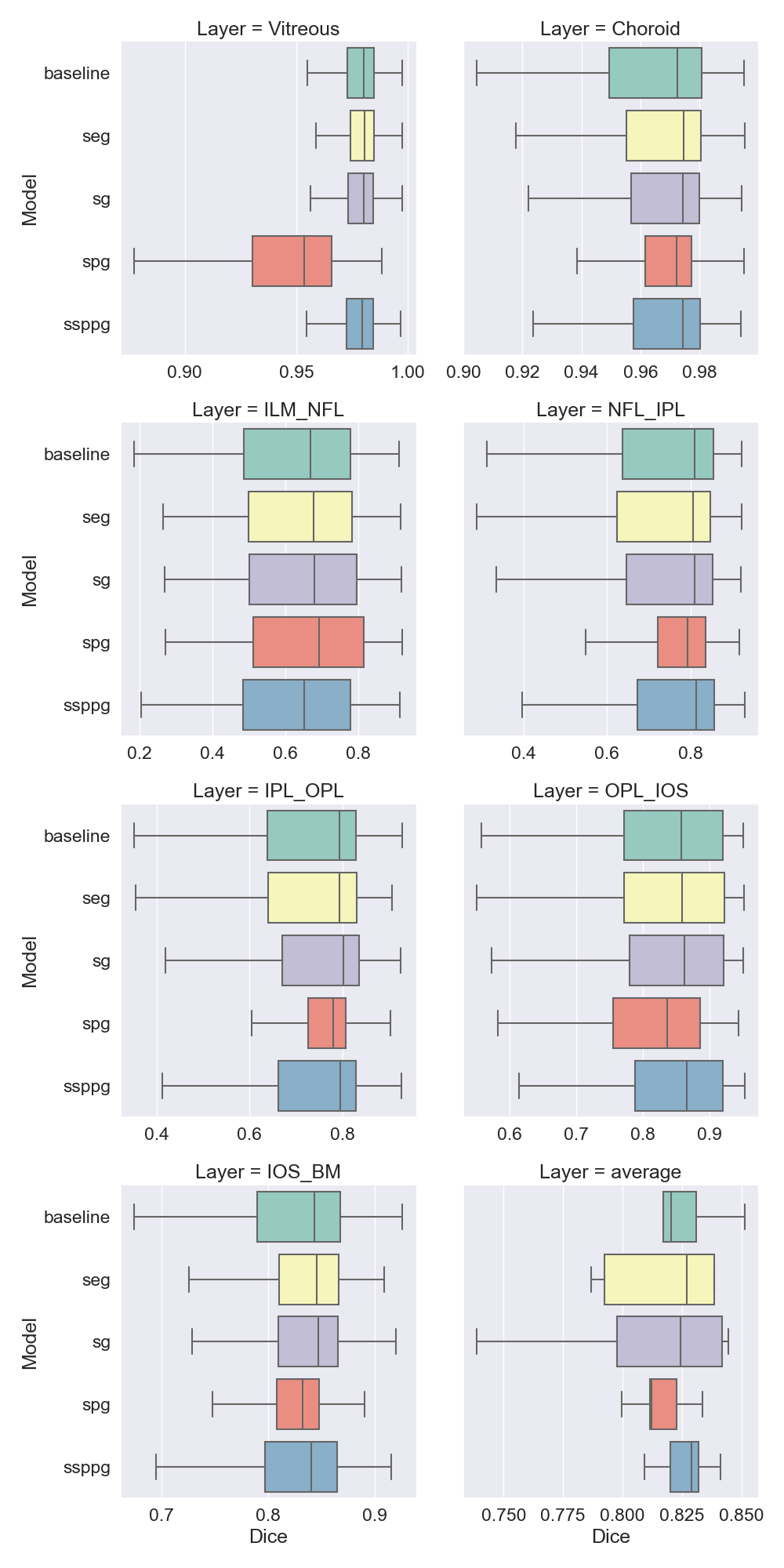}
\caption{Box plots of Dice score for 7 retinal regions averaged among 9 training folds. Each subplot represents the distribution of dice scores for each retinal region for all models, and the last subplot represents the averaged dice scores among all training folds. The median dice are marked within each box, and the whisker is calculated as 1.5IQR(Interquartile Range). The $ssppg$ model has the best average dice score among all models of 0.8231.}
\label{fig:boxplot}
\end{figure}

\begin{figure*}[htb]
\centering
\includegraphics[width=0.9\textwidth]{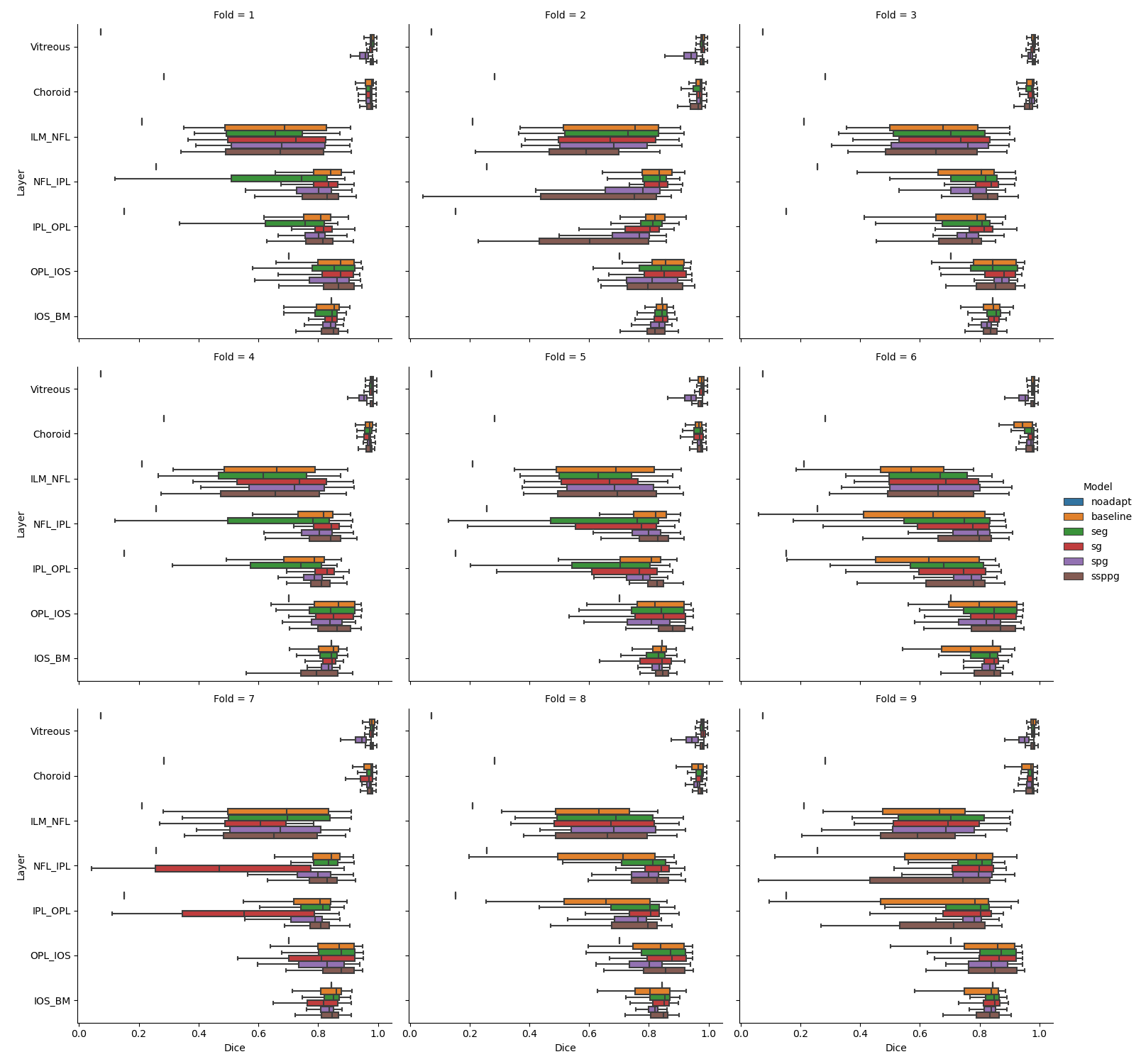}
\caption{Box plots of Dice score for 7 retinal regions for each training fold. The dice score of $noadapt$ model is constant for all folds shown as a vertical line in each subplot. The median dice are marked within each box, and the whisker is calculated as 1.5IQR(Interquartile Range). The $ssppg$ model has the best dice score in folds 5 \& 6 of 0.8413 and 0.8093, respectively. The $sg$ model has the best dice score in folds 1,3,4, and 8 of 0.8438, 0.8419, 0.8446 and 0.8312, respectively. The $seg$ model has the best dice score in folds 7 and 9 of 0.8385 and 0.8300, respectively. The $baseline$ model has the best dice score in fold 2 of 0.8515.}
\label{fig:boxplot_fold}
\end{figure*}

To evaluate the effect of each proposed component in the domain adaptation framework. We performed ablation experiments with 5 different combinations of loss functions, including the original CycleGAN network as our baseline. The other 4 networks add extra constraints onto the baseline network, which are semantic consistency($seg$), semantic consistency + perceptual($sp$), semantic consistency + perceptual + gradient consistency($spg$), and semantic consistency + ssim + psnr + perceptual + gradient consistency($ssppg$). The $seg$ model is equivalent to the method proposed by Hoffman et al.~\citep{Hoffman2018CyCADA:Adaptation}. The $spg$ model is functionally similar to the approach of Li et al.~\citep{Li2019Semantic-awareAdaption}. The SSIM and PSNR loss were combined for experiments as they both evaluate images at the intensity distribution level. The inference results are directly adopted for calculating the dice score as $1 - L_{dice}$ as mentioned in Eq. (1). Figures \ref{fig:boxplot} and \ref{fig:boxplot_fold} show the dice score for all 7 retinal regions of each adaptation model averaged across all 9 training folds as well as in each fold accordingly. The $ssppg$ model has the best dice score in $NFL\_IPL$, $IPL\_OPL$, $OPL\_IOS$ regions, $sg$ has the best dice score in $IOS\_BM$ region, and $spg$ has the best dice score in $ILM\_NFL$ region. Overall, the $ssppg$ model has the optimal average dice score of 0.8231. Furthermore, all 5 adaptation models outperform the one without the domain adaptation. The segmentation performance of detection of vitreous and choroid is similar for all 5 models. 

\begin{figure}[htb]
\centering
\includegraphics[width=0.45\textwidth]{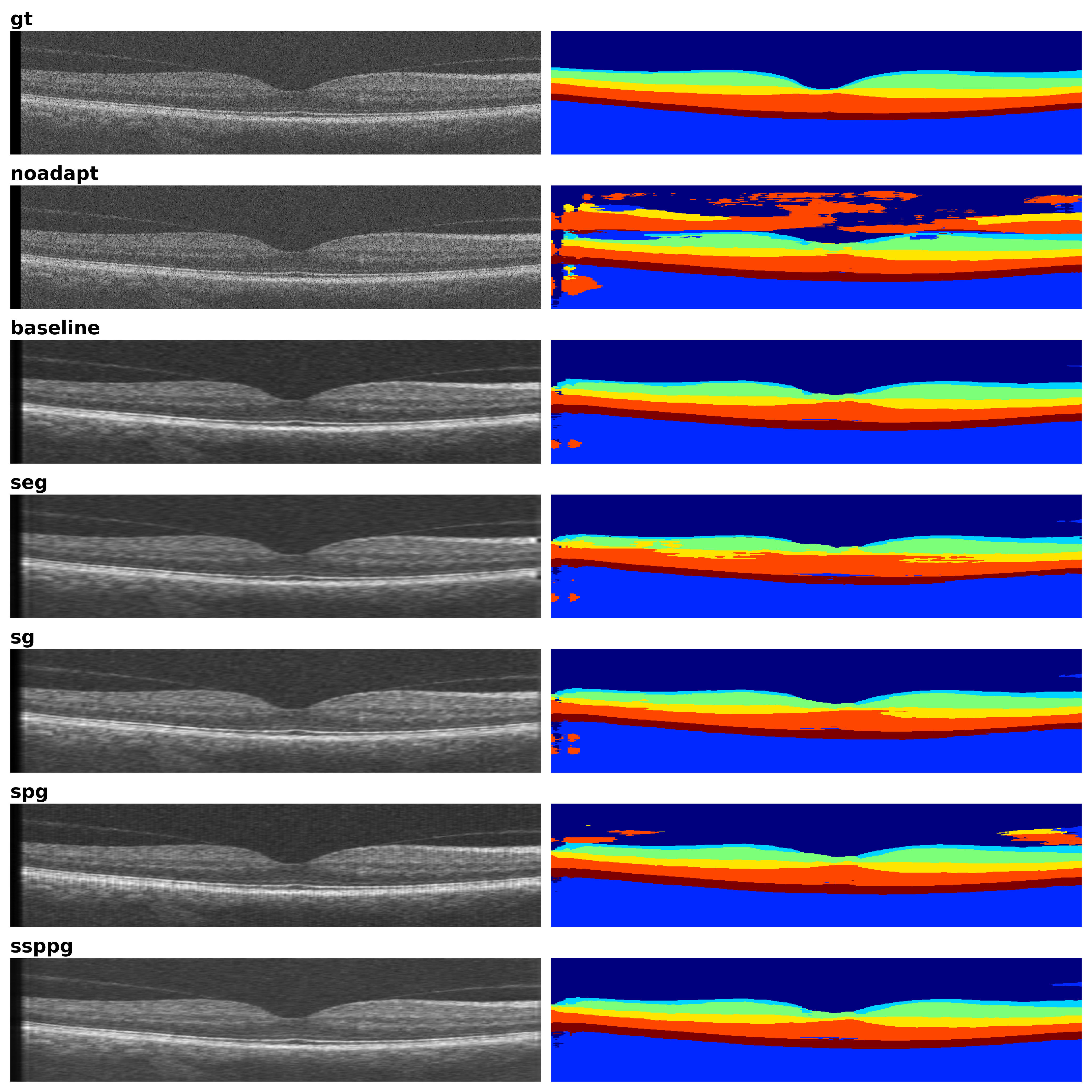}
\caption{Sample results of Bscan near central foveal region. Each row represents the adapted image generated by a specific model and its corresponding segmentation generated by the segmentation network.}
\label{fig:foveal}
\end{figure}

% \begin{figure}[htb]
% \centering
% \includegraphics[width=0.45\textwidth]{photos/montage/artefact.png}
% \label{fig:artefact}
% \caption{Bscan with shadowing artefact. Each row represents the adapted image generated by a specific model and its corresponding segmentation generated by the segmentation network.}
% \end{figure}

\begin{figure}[htb]
\centering
\includegraphics[width=0.45\textwidth]{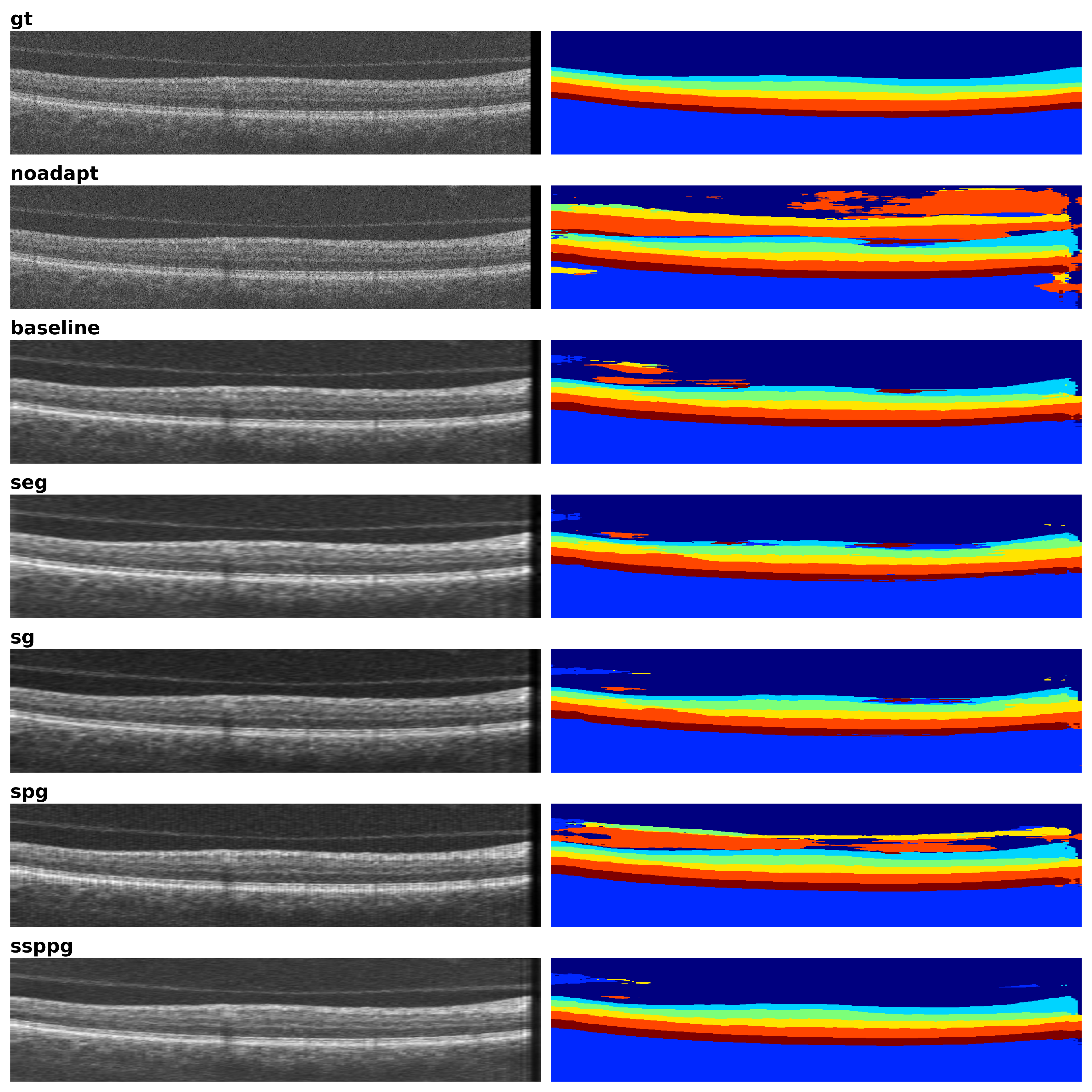}
\caption{Sample results of adapted images with hallucination. Each row represents the adapted image generated by a specific model and its corresponding segmentation generated by the segmentation network.}
\label{fig:hallucination}
\end{figure}

\begin{figure}[htb]
\centering
\includegraphics[width=0.45\textwidth]{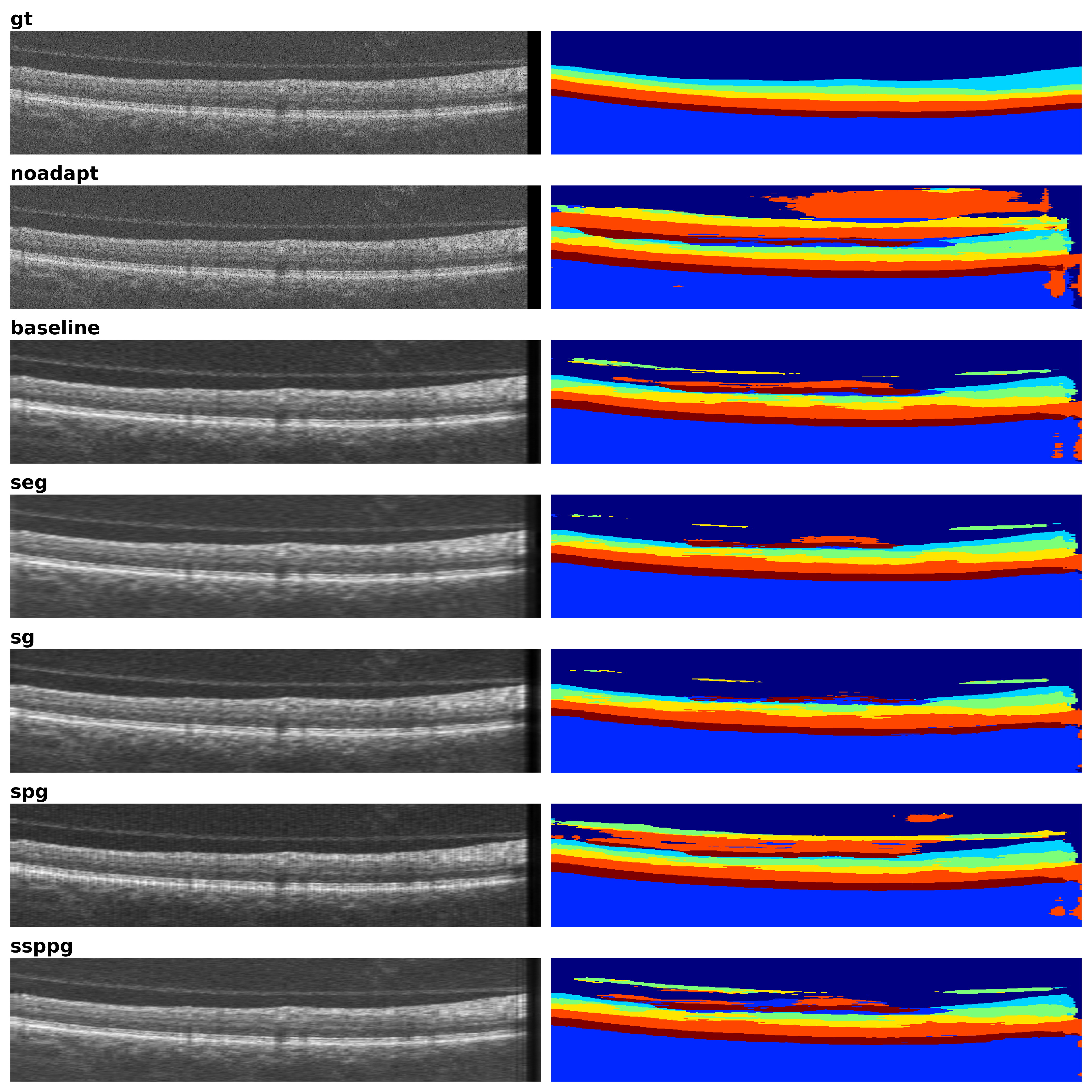}
\caption{Sample results of failed case where the upper retinal layers cannot be delineated. Each row represents the adapted image generated by a specific model and its corresponding segmentation generated by the segmentation network.}
\label{fig:failed}
\end{figure}

We applied statistical analysis to compare the segmentation performance of the 5 adaptation models. The one-way Analysis of Variance (ANOVA) with a post-hoc Tukey Honestly Significant Difference (HSD) test is used for pairwise evaluation of means of 5 models. Similar to Figures \ref{fig:boxplot} and \ref{fig:boxplot_fold}, we calculated the P-value in terms of layers and training folds, which are shown in Tables \ref{tbl:anova_fold} \& \ref{tbl:anova_layer}. Table \ref{tbl:anova_fold} shows the result of the overall mean segmentation performance for each of the fold. The segmentation performance between $baseline$ and $seg$ models, plus $sg$ and $spg$ models are statistically significant by more than half of the training folds, while the $seg$ and $sg$ models plus $spg$ and $ssppg$ models pairs reveals the opposite. Combined with Figure \ref{fig:boxplot}, Table \ref{tbl:anova_layer} shows the results for each of the retinal layers that are averaged across all folds. The $seg$ model statistically improves the segmentation performance in $Vitreous$ and $IOS\_BM$ regions compared to the $baseline$ model. $seg$ and $sg$ models are statistically identical. $spg$ model has statistically better performance than $sg$ in $IPL\_OPL$ and $Choroid$ regions, but worse in $OPL\_IOS$, $IOS\_BM$ and $Vitreous$. $ssppg$ model outperforms $spg$ model in $IPL\_OPL$ and $OPL\_IOS$ regions, but worse in $ILM\_NFL$ region. 

\begin{table*}[htb]
\centering
\begin{tabular}{|c|c|c|c|c|}
\hline
\diagbox{Fold}{P}{Models} & $baseline-seg$ & $seg-sg$ & $sg-spg$ & $spg-ssppg$\\
\hline
1 & \textbf{0.000} & \textbf{0.000} & 0.006 & 0.478\\
\hline
2 & 0.066 & 0.145 & \textbf{0.001} & \textbf{0.001}\\
\hline
3 & 0.908 & 0.224 & 0.908 & 0.283\\
\hline
4 & \textbf{0.015} & \textbf{0.000} & \textbf{0.029} & 0.835\\
\hline
5 & \textbf{0.000} & 0.423 & 0.124 & \textbf{0.000}\\
\hline
6 & \textbf{0.000} & 0.423 & 0.124 & \textbf{0.000}\\
\hline
7 & 0.927 & \textbf{0.000} & \textbf{0.000} & \textbf{0.017}\\
\hline
8 & 0.927 & \textbf{0.000} & \textbf{0.000} & \textbf{0.017}\\
\hline
9 & \textbf{0.000} & 0.839 & 0.258 & \textbf{0.003}\\
\hline
\end{tabular}
\caption{Statistical analysis of the ablation study to evaluate the overall mean segmentation performance among different adaptation models for each of the folds. ANOVA with post-hoc Tukey HSD tests was performed for 4 pairs of adaptation models in each training fold. The significant level is set to 0.05 and the P values below it are highlighted. 6 out of 9 folds show significant statistical difference between $baseline$ and $seg$ models, 6 out of 9 folds show no variations between $seg$ and $sg$ models, 5 out of 9 folds show the explicit difference between $sg$ and $spg$ models, 5 out of 9 folds show no statistical variations between $spg$ and $ssppg$ models.}
\label{tbl:anova_fold}
\end{table*}

\begin{table*}[htb]
\centering
\begin{tabular}{|c|c|c|c|c|}
\hline
\diagbox{Layer}{P}{Models} & $baseline-seg$ & $seg-sg$ & $sg-spg$ & $spg-ssppg$\\
\hline
$ILM\_NFL$ & 0.354 & 0.525 & 0.525 & \textbf{0.000}\\
\hline
$NFL\_IPL$ & 1.000 & 1.000 & 0.101 & 1.000\\
\hline
$IPL\_OPL$ & 1.000 & 0.320 & \textbf{0.028} & \textbf{0.003}\\
\hline
$OPL\_IOS$ & 1.000 & 1.000 & \textbf{0.000} & \textbf{0.000}\\
\hline
$IOS\_BM$ & \textbf{0.000} & 1.000 & \textbf{0.000} & 0.068\\
\hline
$Vitreous$ & 0.251 & 0.251 & \textbf{0.000} & \textbf{0.000}\\
\hline
$Choroid$ & \textbf{0.009} & 0.299 & \textbf{0.000} & 0.245\\
\hline
\end{tabular}
\caption{Statistical analysis of the ablation study to evaluate the overall segmentation performance for each of the retinal layers among different adaptation models, averaged across all the folds. ANOVA with post-hoc Tukey HSD tests was performed for 4 pairs of adaptation models in each retinal region. The significant level is set to 0.05 and the P values below it are highlighted. The difference in segmentation performance between $baseline$ and $seg$ are significant in $IOS\_BM$ and $Choroid$. $seg$ and $sg$ model show no statistical difference in all retinal regions. $spg$ model is significantly different with $sg$ model in all regions except $ILM\_NFL$ and $NFL\_IPL$. $spg$ and $ssppg$ models have a significant difference in $ILM\_NFL$, $IPL\_OPL$, $OPL\_IOS$ and $Vitreous$. }
\label{tbl:anova_layer}
\end{table*}

Figures \ref{fig:foveal} \& \ref{fig:hallucination} show the successful examples of adapted images and corresponding segmentation. The first column shows the original UKB image, image without adaptation, adapted images from $baseline$, $seg$, $sg$, $spg$ and $ssppg$, respectively. The second column shows the corresponding segmentation for each experiment set. The segmentation results are obtained directly from the model's output without post-processing steps. Figure \ref{fig:foveal} shows a example Bscan near the central foveal region. We observe a better segmentation performance near the foveal pit for $ssppg$ model compared to other models, and $sspp$g model is more robust to produce minimal noisy segmentation. Figure \ref{fig:hallucination} 
demonstrates the effect of possible hallucination of retinal detachment potentially due to background noise. The $seg$, $sg$, and $spg$ models all failed to produce clean segmentation of the upper retinal layers(e.g. $ILM\_NFL$) and Vitreous, but the $ssppg$ model mitigates such issue even near the right-hand side region, where the image frame is shifted due to motion correction. Figure \ref{fig:failed} shows the failed circumstance where the $ILM\_NFL$ and $NFL\_IPL$ layers cannot be properly segmented by all models. It might be caused by the central shadowing artifact of the retina as well as the noise in the Vitreous region. 

We also evaluated the efficiency of the model via total training time under the same early stopping criteria. As mentioned in Section \ref{network_adapt}, the training stops when the validation semantic consistency loss of the UKB dataset converged. Table \ref{tbl:training_time} demonstrates the training time in hours for each model in each training fold. We observed that both $spg$ and $ssppg$ models are trained significantly faster than other models, benefiting from the perceptual loss without 
affecting segmentation qualities. 
\begin{table*}[htb]
\centering
\begin{tabular}{|c|c|c|c|c|c|}
\hline
\diagbox{Fold}{Training time(hours)}{Model} & $baseline$ & $seg$ & $sg$ & $spg$ & $ssppg$\\
\hline
1 & 51.12 & 60.96 & 64.08 & \textbf{52.08} & 76.32\\
\hline
2 & 55.92 & 61.92 & 47.04 & 48.00 & \textbf{35.04}\\
\hline
3 & 70.08 & 54.00 & 53.04 & \textbf{25.92} & 43.92\\
\hline
4 & 40.08 & 49.92 & 70.08 & 33.12 & 34.08\\
\hline
5 & 70.08 & 70.08 & 51.12 & \textbf{42.96} & 54.00\\
\hline
6 & 34.08 & 70.08 & 55.92 & 41.04 & \textbf{33.12}\\
\hline
7 & 54.00 & 61.92 & 66.96 & \textbf{36.00} & 39.12\\
\hline
8 & 54.00 & 61.92 & 39.12 & \textbf{24.00} & 64.08\\
\hline
9 & 54.00 & 53.04 & 54.00 & 55.92 & \textbf{39.12}\\
\hline
average & 53.71 & 60.43 & 55.71 & \textbf{39.89} & 46.53\\
\hline
\end{tabular}
\caption{Total training time in hours for each model among 9 training folds. The time is calculated based on early stopping criteria where the validation loss of Topcon data segmentation converges. The smallest training time for each fold is highlighted. Both $spg$ and $ssppg$ models significantly reduce the training time compared to the other 3 models.}
\label{tbl:training_time}
\end{table*}

\subsection{Image metrics}\label{results_metrics}

The quantitative measure of the FID score was applied to the original and the adapted Zeiss dataset. We wanted to evaluate the image similarity of the synthesized Zeiss dataset adapted from the Topcon dataset. Due to the limited number of validation data(128 Bscans), we computed the Fréchet distance for 4 dimensionalities of feature maps from Inception V3, which are 64 from the first max pooling layer, 192 from the second max pooling layer before the Inception module A, 768 from the output of the Inception module A, and 2048 from the final average pooling layer~\citep{MaximilianSeitzer2020Pytorch-fid:PyTorch}. The results are shown in Table \ref{tbl:fid}, the $spg$ model and $ssppg$ model obtain the lowest and the second lowest FID scores among all adaptation models. And the FID scores of all adaptation models are significantly smaller than the one without adaptation. Table \ref{tbl:kid} illustrates the evaluation using KID score. We set $\gamma$ to be 2048 matching the output features of the InceptionV3 model, the bias constant $c$ to be 1 and degree of polynomial to be 3. Results show that the $ssppg$ model has the lowest average KID score compared to other 5 adaptation models in 7 training folds, and $sg$ model has the best KID score in the rest of 2 folds. Similar to results in Table \ref{tbl:fid}, all the adaptation models have much better performance than $noadapt$ model.

\begin{table*}[htb]
\centering
\begin{tabular}{|c|c|c|c|c|c|c|}
\hline
\diagbox{Dim}{FID}{Model} & $noadapt$ & $baseline$ & $seg$ & $sg$ & $spg$ & $ssppg$ \\
\hline
$64$ & 151304.575 & 23.196 & 23.697 & 23.285 & \textbf{20.950} & 23.639 \\
\hline
$192$ & 345710.933 & 41.908 & 42.629 & 42.106 & \textbf{28.957} & 36.922 \\
\hline
$768$ & 450.540 & 1.949 & 1.996 & 1.985 & \textbf{1.670} & 1.828 \\
\hline
$2048$ & 2175.558 & 245.028 & 250.993 & 250.230 & \textbf{198.623} & 218.106\\
\hline
\end{tabular}
\caption{FID score for each dimensionality of the feature map. The lowest FID scores among all models are highlighted, which indicates better image similarities to the original image. The scores for each model is averaged among all 9 training folds except for $noadapt$. The $spg$ and $ssppg$ models have the lowest and second lowest FID scores in all dimensionalities.}
\label{tbl:fid}
\end{table*}

\begin{table*}[htb]
\centering
\begin{tabular}{|c|c|c|c|c|c|c|}
\hline
\diagbox{Fold}{KID}{Model} & $noadapt$ & $baseline$ & $seg$ & $sg$ & $spg$ & $ssppg$\\
\hline
1 & 0.9659 \textpm  0.0375 & 0.5559 \textpm  0.2082 & 0.0456 \textpm  0.0614 & \textbf{0.0060 \textpm  0.0174} & 0.0234 \textpm  0.0383 & 0.0088 \textpm  0.0289 \\
\hline
2 & 0.9659 \textpm  0.0375 & 0.3541 \textpm  0.2041 & 0.0429 \textpm  0.0586 & 0.0082 \textpm  0.0214 & 0.0147 \textpm  0.0294 & \textbf{0.0052 \textpm  0.0189} \\
\hline
3 & 0.9659 \textpm  0.0375 & 0.2505 \textpm  0.1716 & 0.0498 \textpm  0.0676 & 0.0069 \textpm  0.0219 & 0.0121 \textpm  0.0270 & \textbf{0.0052 \textpm  0.0105} \\
\hline
4 & 0.9659 \textpm  0.0375 & 0.1876 \textpm  0.1691 & 0.0278 \textpm  0.0495 & \textbf{0.0056 \textpm  0.0154} & 0.0164 \textpm  0.0376 & 0.0082 \textpm  0.0218 \\
\hline
5 & 0.9659 \textpm  0.0375 & 0.1244 \textpm  0.1208 & 0.0220 \textpm  0.0386 & 0.0035 \textpm  0.0136 & 0.0109 \textpm  0.0273 & \textbf{0.0027 \textpm  0.0125} \\
\hline
6 & 0.9659 \textpm  0.0375 & 0.1134 \textpm  0.1165 & 0.0369 \textpm  0.0529 & 0.0053 \textpm  0.0180 & 0.0086 \textpm  0.0200 & \textbf{0.0048 \textpm  0.0146} \\
\hline
7 & 0.9659 \textpm  0.0375 & 0.0750 \textpm  0.0936 & 0.0246 \textpm  0.0473 & 0.0048 \textpm  0.0146 & 0.0180 \textpm  0.0346 & \textbf{0.0046 \textpm  0.0189} \\
\hline
8 & 0.9659 \textpm  0.0375 & 0.0704 \textpm  0.0903 & 0.0270 \textpm  0.0501 & 0.0042 \textpm  0.0147 & 0.0143 \textpm  0.0347 & \textbf{0.0038 \textpm  0.0137} \\
\hline
9 & 0.9659 \textpm  0.0375 & 0.0584 \textpm  0.0766 & 0.0231 \textpm  0.0457 & 0.0064 \textpm  0.0249 & 0.0111 \textpm  0.0242 & \textbf{0.0041 \textpm  0.0111} \\
\hline
Average & 0.9659 \textpm  0.0375 & 0.1989 \textpm  0.1390 & 0.0333 \textpm  0.0524 & 0.0057 \textpm  0.0180 & 0.0144 \textpm  0.0304 & \textbf{0.0053 \textpm  0.0167} \\
\hline
\end{tabular}
\caption{KID score of 5 models with all 9 folds. The final row shows the average KID score among all folds. The smallest KID score for each fold is highlighted. The score for $noadapt$ is fixed for all folds since there is no variation among different folds. The $ssppg$ model has the smallest averaged mean and standard deviation of 0.0053 \textpm  0.0167 compared to the other models}
\label{tbl:kid}
\end{table*}

\begin{table*}[htb]
\centering
\begin{tabular}{|c|c|c|c|c|c|c|c|c|}
\hline
\diagbox{Model}{Dice}{Layer} & $ILM\_NFL$ & $NFL\_IPL$ & $IPL\_OPL$ & $OPL\_IOS$ & $IOS\_BM$ & $Vitreous$ & $Choroid$ & Average \\
\hline
$ssppg\_adapt$ & 0.5065 & 0.6177 & 0.6694 & 0.5237 & 0.7049 & 0.9082 & 0.7350 & 0.7476\\
\hline
$Direct$ & 0.9597 & 0.9652 & 0.9309 & 0.9171 & 0.9375 & 0.9987 & 0.9997 & 0.9560\\
\hline
\end{tabular}
\caption{Dice Score for two extra experiments. $ssppg\_adapt$ indicates the GAN-based transfer learning using only adapted UKB images. $Direct$ represents the direct training using original UKB image and label pairs. The average dice score is calculated for 7 retinal regions. The $ssppg\_adapt$ model does not produce comparable results to our adaptation models. The $Direct$ model has better performance, which uses the true image and label pairs.}
\label{tbl:dice_ext}
\end{table*}

\section*{Discussion}\label{discussion}
As shown in Section \ref{results_seg}, the segmentation network achieves the state-of-art performance on all retinal layers. It sets the upper limit for the segmentation results on the adapted images. The $ssppg$ model improves the overall dice performance by an average of 46.2\% with respect to no adaptation, and its performance reaches 87.4\% of the first-stage segmentation model in terms of average Dice score. From Figures \ref{fig:boxplot} and \ref{fig:boxplot_fold}, all the adaptation models show similar performance for $Vitreous$ and $Choroid$ layers.  

As shown in Figures \ref{fig:foveal}-\ref{fig:failed}, the segmentation network is very sensitive to the domain shifts of the images. The delineation of retinal layers are difficult since the semantic labels are allocated based on both local gradient changes and global feature distributions. The hallucination shown in Figures \ref{fig:hallucination} \& \ref{fig:failed} is mainly caused by the lateral shifts and unwanted background noise of the original image, which can be induced during data acquisition or pre-processing steps like speckle noise due to scattering and interference of coherent light and motion corrections. For the $seg$ model, the undefined shifted region may be mistakenly identified as a topological feature of the retina. The gradient loss significantly mitigates such a problem by enforcing the learning of the ``true'' local boundaries, which act as a more generalized semantic constraint using the gradient map. However, as mentioned in Section \ref{Intro}, the GAN network usually suffers from long training time and complicated parameter tuning. As shown in Table \ref{tbl:training_time}, both $spg$ and $ssppg$ models converge significantly faster than the other models due to the perceptual loss. Unlike the traditional style-transfer networks, the perceptual loss is calculated using a pre-trained classification network(e.g. VGG16) with weights frozen, which significantly speeds up the training process while maintaining the high-level texture of the original images. However, the classification network pre-trained on ImageNet may not precisely extract high-level features from unseen images like retinal OCT. The differentiation of the adjacent retinal layers relies more on pixel-level details than global features. Such noise may represent propagation of the high-level deviations. Therefore, both SSIM and PSNR criterion contribute to noise reduction throughout the image. The discrepancy of two intensity distributions can be reduced via matching both the noise distributions and the pixel information. Even though the time cost increases compared to the $spg$ model, the improvement of the segmentation performance is worth the trade-off. 

% \begin{table*}[htb]
% \centering
% \begin{tabular}{|c|c|c|c|c|c|c|c|c|}
% \hline
% \diagbox{Model}{Ratio(\%)}{Layer} & $ILM\_NFL$ & $NFL\_IPL$ & $IPL\_OPL$ & $OPL\_IOS$ & $IOS\_BM$ & $Vitreous$ & $Choroid$ & Average \\
% \hline
% noadapt & 68.15 & 74.19 & 71.87 & 79.20 & 78.96 & 93.69 & 98.73 & 80.68\\
% \hline
% baseline & 73.55 & 77.18 & 79.77 & 86.14 & 81.84 & \textbf{99.13} & 100.15 & 85.39\\
% \hline
% seg/Hoffman et al. & 72.84 & 77.82 & 81.45 & 87.14 & 82.48 & 98.94 & 99.92 & 85.80\\
% \hline
% sg & 77.53 & 81.79 & 82.95 & 88.72 & \textbf{84.70} & 98.80 & 100.47 & 87.85\\
% \hline
% spg/Li et al. & 78.19 & 81.82 & 82.58 & \textbf{89.38} & 84.24 & 98.24 & \textbf{100.50} & 87.85\\
% \hline
% ssppg & \textbf{80.74} & \textbf{85.80} & \textbf{84.86} & 89.22 & 84.62 & 98.07 & 100.35 & \textbf{89.09}\\
% \hline
% % ssppg\_adapt & 0.9482 & 0.7350 & 0.7567 & 0.7956 & 0.6694 & 0.5237 & 0.8049 & 0.7476\\
% % \hline
% \end{tabular}
% \caption{Change of Mean Dice Score (in terms of percentage point) with respect to the first-stage segmentation model. The best percentage among all models are highlighted. The $ssppg$ model achieves 89.09\% of the performance of the first-stage segmentation network, which increases by 8.41\% from the model without adaptation, and 3.7\% from the baseline model.}
% \label{tbl:dice_ratio}
% \end{table*}

While the segmentation of the retinal OCT is our main interest, we also investigated the relationship between the segmentation performance and the image quality characteristics. Specifically, the adapted images can be re-usable if the segmentation-favoured domain is close to the actual target domain. As illustrated in Section \ref{results_metrics}, the images adapted by $spg$ and $ssppg$ models have lower FID and KID scores than the adaptation models. Both the FID and KID scores tend to decrease along with the complexities of the loss functions. It inferred that the adapted images with better segmentation can lead to higher similarities of image domains, i.e. the segmentation model can produce higher quality semantic labels from a latent space similar to the original image distribution. 

We also performed a follow-up experiment using the best-performance $ssppg$ model to re-train the segmentation network with only the adapted images, which can be regarded as a GAN-based transfer learning approach. We used the same training settings as mentioned in Section \ref{network_seg}, but both the training and validation data were the adapted UKB images, i.e. the synthesized UKB images adapted from the Zeiss dataset. The segmentation performance was evaluated by directly feeding the original UKB testing dataset into the re-trained segmentation network. The number of training data is vastly increased compared to the ones used in the adaptation network, but the segmentation performance is not comparable to any of the previous models. The results are shown in Table \ref{tbl:dice_ext} named $ssppg\_adapt$. It revealed that the extra transfer learning can worsen the segmentation performance, and the adapted images can still be distinguishable from the original images. A possible remedy for this experiment would be including the original UKB images along with its pseudo-labels generated by its adapted images adapted from the Zeiss dataset, thus the network can learn from the original UKB data while the pseudo-labels can also contribute to the model training. Nevertheless, the dice scores may be more suitable to amplify the differences than the commonly used image quality metrics, and it is more of clinical usage in various analytical applications.

\section{Limitation and Future Work}
% Due to time constraints and complexity of the experiment design, 
There are several limitations to our proposed two-stage pipeline. The corresponding remedies and possible future work are elaborated on in this section.

% \begin{itemize}
First, We did not quantitatively evaluate the performance of the fine-tuning strategies. We experienced high failure rates of training when intensity-level transformations were involved, such as noise injection, interpolations due to deformation, modifications of luminance, contrast, etc. Similar issues occurred when we used optimizers with momenta like Adam and SGD. The choices of proper learning rate was crucial for a GAN-based network, especially the integration of several losses were involved. Multiple researches showed that lower learning rates shall always be considered, but the network can still fail to converge even with smaller learning rates and longer training time~\citep{Yazc2019THETRAINING}. The TTUR concept was rigorously evaluated by Heusel et al., which benefited more practically for testing and learning rate tuning. The soft labels were proposed in WassersteinGAN, which also benefited the LeakyReLU activation in the discriminator module. We experienced a relatively faster convergence for the baseline model using soft labels, but the segmentation performance were roughly unchanged. As all adaptation models were tested using the same strategies, it will not interfere with the final evaluation results. We also performed several experiments adjusting the weighting parameters($\lambda$) of the loss functions both statically and dynamically. Due to the constraints of the resources, we tried few fixed combinations of parameters, also linearly increasing the weight of the certain loss functions as training carries out. Unfortunately, these adjustment did not lead to better performance or even causes structural collapses of the images. Systematic parameter optimization could potentially be performed in future studies like grid search to obtain the optimal settings. 

The Topcon dataset only contain OCT images from healthy patients, where the segmentation model was trained the Zeiss dataset with several pathological cases. To avoid false positives of fluid, we ignored the fluid channel from the output of the Softmax activation layer. An extended study can be performed mainly on pathological data, where the performance of fluid segmentation can be evaluated separately. Due to resource constraints, we did not perform experiments on the public external dataset, which required a decent number of image and label pairs for both training and validation. However, the structural appearance of SD-OCT images acquired from various devices is visually similar. The successful adaptation with a limited number of UKB images proves the robustness and adaptability of our model. Additionally, the number of retinal layers can be further enriched, which will apply better constraints to the adaptation. Empirically, more semantic constraints shall further improve the performance of the network, providing that our segmentation network is trained successfully. 

We investigated the segmentation performance using only Topcon dataset with ground-truth labels used for evaluation of our GAN network. Table \ref{tbl:dice_ext} shows the training results named $Direct$. It outperforms the $ssppg$ model as expected. It directly uses the true UKB data and labels. However, our proposed method aimed to generate segmentation when the data in the target domain has no ground-truth labels available. Without such a prerequisite, an extra experiment can be delivered that we can add segmentation constraints using UKB labels too, which shall produce considerably better results compared to the current $ssppg$ model. 

The segmentation performance was evaluated among raw outputs of different models. A series of post-processing methods can further improve the performance of the models, such as keeping the largest connected components of the labels, hole fillings, proper erosion and dilation, etc. However, such labels still need to be manually corrected for analytical measurements. The comparison of the raw inference results can better reflect the performance of the models.

\section{Conclusion}
In this study, we proposed a novel two-stage segmentation-guided domain adaptation network based on CycleGAN architecture to achieve effective data harmonization for multi-site OCT data. The proposed approach significantly improves the segmentation results for images of different domains. Once the adaptor module is trained for specific OCT acquisition devices, the preliminary results of good qualities can be produced in real-time to speed up the process of manual corrections. Future work shall be focused on the adaptation of pathological OCT images where the presence of fluid causes explicit deformation or even destruction of the retinal layers. Better performance can be obtained with a follow-up of post-processing steps.

\section{Funding and Acknowledgement}
This study was funded by the National Sciences and Engineering
Research Council of Canada; Canadian Institutes of Health Research, Compute Canada, Wake Forest University School of
Medicine Startup Funding, and Precision Imaging Beacon, University of
Nottingham. The sponsor or funding organization had no role in the
design or conduct of this research.

\section{Declaration of competing interest}
There is no conflict of interest declared from all co-authors

\newpage
% \section*{References}
% \bibliographystyle{./bibliography/IEEEtranS}
\bibliographystyle{plainnat}
\bibliography{./bibliography/CycleGAN_combined.bib}

\end{document}